\documentclass{article}

\usepackage{arxiv}

\usepackage[utf8]{inputenc} % allow utf-8 input
\usepackage[T1]{fontenc}    % use 8-bit T1 fonts
\usepackage{hyperref}       % hyperlinks
\usepackage{url}            % simple URL typesetting
\usepackage{booktabs}       % professional-quality tables
\usepackage{amsfonts}       % blackboard math symbols
\usepackage{nicefrac}       % compact symbols for 1/2, etc.
\usepackage{microtype}      % microtypography
\usepackage{lipsum}
\usepackage{graphicx}
\graphicspath{ {./images/} }
\usepackage{amssymb}   % for math
\usepackage{amsmath}
\usepackage{longtable}
\usepackage{mathtools} 
\usepackage{tikz}
\usepackage{lscape}
\usepackage{array}
\usepackage{cite}

\newcommand{\scae}{^{(1)}}
\newcommand{\bu}{^{(0)}}
\newcommand{\im}{\text{Im}}
\newcommand{\sub}[1]{_\text{#1}}
\newcommand{\tr}{\text{tr}}
\newcommand{\intd}[3]{\int\limits_{#1}^{#2}\,\text{d}#3\,}
\newcommand{\intdz}[3]{\int_{#1}^{#2}\,\text{d}^2#3\,}
\newcommand{\ev}[1]{\langle#1\rangle}
\newcommand{\kb}{k\sub{B}}
\newcommand{\tc}{\text{c}}
\newcommand{\sca}{\mathbf{G}^{(1)}}
\newcommand{\ti}{\text{i}}
\newcommand{\cev}[1]{\reflectbox{\ensuremath{\vec{\reflectbox{\ensuremath{#1}}}}}}

\newcommand{\scacurlt}[2]{\nabla\times\sca(#1,#2)\times\cev{\nabla}\Big\vert_{#2=#1}}
\newcommand{\mats}{\sum\limits_{m=0}^{\infty}{\vphantom{\sum}}'}
\newcommand{\fe}{\mathbf{e}}
\newcommand{\kapp}{\kappa^\perp}
\newcommand{\kapps}{\kapp{\vphantom{\kappa}}^2}
\newcommand{\kpars}{k^\parallel{\vphantom{k}}^2}

\title{Comparison of theory and experiments on Van der Waals forces in media -- a survey}

\author{
Friedrich Anton Burger\thanks{Current address: Climate and Environmental Physics, Institute of Physics, University of Bern, Sidlerstr. 5, 3012 Bern, Switzerland.}\\
  Institute of Physics,\\ Albert-Ludwigs University of Freiburg,\\ Hermann-Herder-Str. 3, 79104 Freiburg, Germany.\\
 \texttt{friedrich.burger@climate.unibe.ch}\\
 \And
 Robert William Corkery\thanks{
Department of Applied Mathematics, Research School of Physics, The Australian National
University, Canberra, ACT 2601, Australia.
}\\
Applied Physical Chemistry,\\ KTH Royal Institute of Technology,\\ SE 100 44 Stockholm.\\
\texttt{corkery@kth.se} 
\And
 Stefan Yoshi Buhmann \\
  Institute of Physics,\\ Albert-Ludwigs University of Freiburg,\\ Hermann-Herder-Str. 3, 79104 Freiburg, Germany.\\
\And
 Johannes Fiedler\thanks{Centre for Materials Science and Nanotechnology,
Department of Physics, University of Oslo, P. O. Box 1048 Blindern,
NO-0316 Oslo, Norway.} \\
  Institute of Physics,\\ Albert-Ludwigs University of Freiburg,\\ Hermann-Herder-Str. 3, 79104 Freiburg, Germany.
}

\begin{document}
\maketitle
\begin{abstract}
  We present a critical overview comparing theoretical predictions and measurements of Van der Waals dispersion forces in media on the basis of the respective Hamaker constants. To quantify the agreement, we complement the reported experimental errors with those for the theoretical predictions, which are due to uncertainties in the underlying  spectroscopic data. Our main finding is that the theoretical errors are often larger than their experimental counterparts. Within these uncertainties, the comparison confirms the standard Lifshitz theory based on the Abraham electromagnetic stress tensor against the recently suggested alternative account on the basis of the Maxwell stress tensor.
\end{abstract}

% keywords can be removed

\section{Introduction}
The Casimir effect is the mostly attractive force between two dielectric objects~\cite{Casimir482}. Different interpretations for the physical origin of this effect exist, ranging from the radiation pressure of virtual photons on the plates~\cite{Casimir482} or the zero-point energy of the field between the plates~\cite{doi:10.1098/rspa.1963.0025} to the fluctuations of the ground-state quantised electromagnetic field~\cite{PhysRevA.68.033810}. In free space, the Casimir force has been measured with high precision~\cite{PhysRevD.75.077101,Lamoreaux_2004}, confirming the relevance of retardation and thermal field fluctuations. Theoretical approaches to calculating Casimir forces are based on Lifshitz theory which requires a correct modelling and accurate knowledge of the optical response of the interacting materials. This explains the prominence of relatively simple metals and dielectrics in high-precision experiments. The sensitivity of theoretical predictions to the employed dielectric permittivities has become painfully obvious in the recent Drude--plasma debate where the use of the Drude or plamsa models led to strong discrepancies~\cite{Lamoreaux_2004,PhysRevA.90.062115}.

The characterisation of dispersion forces across an intervening liquid is more challenging both experimentally and theoretically. Measurements typically focus on distance regimes where retardation can be disregarded and one speaks of the Van der Waals interaction whose strength is determined by the Hamaker constant. These investigations are motivated in part by their obvious relevance to colloidal and biological systems and in part by the intriguing possibility of making dispersion forces repulsive by the introduction of the intervening medium~\cite{Munday2009,ZWOL,Tang2017}. The repulsive dispersion force between an optically thick body and an optically thin one across a liquid of intermediate optical strength is akin to the upwards buoyancy force experienced by a gas bubble under water. A similar setup of two objects in a medium is also central to the critical Casimir effect~\cite{Hertlein2008}. 

In addition to the additional uncertainty in the theoretical description brought about by the often poorly characterised properties of the medium, theoretical predictions of Van der Waals forces in liquids have recently been questioned on much more fundamental grounds. While Lifshitz theory for Van der Waals forces in free space is conventionally generalised to media on the basis of the Abraham stress tensor~\cite{Abraham2009,Minkowski1910} for the electromagnetic field~\cite{Dzyaloshinskii_1961}, Raabe and Welsch had advocated the use of the Maxwell stress tensor as it could be shown to be equivalent to the Lorentz force and was able to predict a Casimir-induced pressure gradient within the intervening medium~\cite{PhysRevA.71.013814}. This idea had been challenged by several authors~\cite{Pitaevskii2006,PhysRevA.79.027801,robinson1975}. Some of us have recently given theoretical arguments in favour of the standard Abraham-based Lifshitz theory by showing its compatibility with microscopic calculations, which are at odds with the Maxwell-based Raabe--Welsch result~\cite{Friedrich}. Note that in a wider context, the momentum of the macroscopic electromagnetic field has been subject to an old debate~\cite{OBUKHOV2003277,RevModPhys.79.1197,Richter_2008,Rubinsztein_Dunlop_2016,kemp2017,Friedrich}. Ultimately, this question can only be settled by experiments, as seems feasible following recent developments increasing the precision of Casimir measurements~\cite{doi:10.1002/cbic.201700327,eafa78c1cce044beb2ff803dcd378cf1}. 

In this manuscript, we consider the Abraham and Maxwell stress tensor approaches to derive two competing expressions for the Casimir force in a planar system. Afterwards, we apply both results to existing experimental setups and compare their predictions with experimental findings. In order to be able to quantify the degree of agreement of each prediction with the experimental data, we estimate the theoretical errors brought about by uncertainties in the optical data on which the Casimir-force calculation is based. The article is structured as follows. In Section~\ref{section:theory}, we present the two competing theoretical predictions for the Casimir force between two parallel plates immersed in a medium on the basis of the two alternative stress tensors and show how their uncertainties derive from those of the optical data. The latter are introduced in Section~\ref{sec:modeling_dielectric_functions} for the materials included in this study. The predicted Casimir forces and their errors are presented and compared with experimental findings in Section~\ref{Sec4}.

\section{The Casimir force for the competing stress tensors}\label{section:theory}

In order to obtain the difference between the Casimir forces based on the different stress tensor approaches, we estimate the force for the simple geometry consisting of two parallel infinite plates separated by a distance for both stress tensors. The Abraham stress tensor for the electromagnetic field is given by~\cite{Friedrich}
\begin{align}
\begin{split}
\mathbf{T}\sub{A}(\mathbf{r},t)=&\mathbf{D}(\mathbf{r},t)\mathbf{E}(\mathbf{r},t)+\mathbf{H}(\mathbf{r},t)\mathbf{B}(\mathbf{r},t)-\frac{1}{2}\Big[\mathbf{D}(\mathbf{r},t)\cdot\mathbf{E}(\mathbf{r},t)+\mathbf{H}(\mathbf{r},t)\cdot\mathbf{B}(\mathbf{r},t)\Big]\mathbb{I}\,,
\end{split}
\label{eq:2_1}
\end{align}
whereas the Maxwell stress tensor reads~\cite{jackson1998classical}
\begin{align}
\begin{split}
\mathbf{T}\sub{M}(\mathbf{r},t)=&\varepsilon_0\mathbf{E}(\mathbf{r},t)\mathbf{E}(\mathbf{r},t)+\frac{1}{\mu_0}\mathbf{B}(\mathbf{r},t)\mathbf{B}(\mathbf{r},t) -\frac{1}{2}\Big[\varepsilon_0\mathbf{E}^2(\mathbf{r},t)+\frac{1}{\mu_0}\mathbf{B}^2(\mathbf{r},t)\Big]\mathbb{I}\,,
\end{split}
\label{eq:2_2}
\end{align}
with electric field $\mathbf{E}(\mathbf{r},t)$, induction field $\mathbf{B}(\mathbf{r},t)$, displacement field
\begin{equation}
    \mathbf{D}(\mathbf{r},t) = \int \mathrm d \tau\,\varepsilon_0 \varepsilon(\tau) \mathbf{E}(\mathbf{r},t-\tau) \, ,
    \label{eq:2_3}
\end{equation}
 magnetic field
\begin{equation}
    \mathbf{H}(\mathbf{r},t) = \int \mathrm d \tau\,\frac{1}{\mu_0 \mu(\tau)} \mathbf{B}(\mathbf{r},t-\tau) \, ,
    \label{eq:2_4}
\end{equation}
and the three-dimensional unit-matrix $\mathbb{I}$. Furthermore, we have introduced the electric permittivity $\varepsilon(\mathbf{r},t)$ (or dielectric function) and the magnetic permeability $\mu(\mathbf{r},t)$. In this study we consider only dielectric materials for which $\mu\equiv1$. To calculate the force density acting on the plates, one must evaluate the quantum-mechanical expectation values of the stress tensors \eqref{eq:2_1} and \eqref{eq:2_2}. Further information can be found in Supporting Information Section S1. The Casimir force per area acting on a slab that is at distance $l$ to another slab (the $z$-axis is chosen perpendicular to the slabs' surfaces) is given by
\begin{equation}
\mathbf{f}=-\fe_z\cdot\ev{\mathbf{T}}\Big\vert_{z=l}=-\ev{\mathbf{T}_{zz}}\Big\vert_{z=l}\fe_z\,,
\label{eq:fd}
\end{equation}
with the quantum-mechanically averaged stress tensor given by Eqs.~\eqref{eq:result_abraham} and \eqref{eq:result_maxwell} in the Appendix. We assume that the plates are close to each other compared to the system’s relevant wavelengths. Introducing the polylogarithmic function
\begin{equation}
    \text{Li}_3(y)=4\intd{0}{\infty}{x}x^2\frac{y e^{-2x}}{1-y e^{-2x}} 
\end{equation}
and defining Hamaker constants for the two theories as
\begin{equation}
    H_{\rm A[M]} = \frac{3\kb T}{2} \mats \frac{1}{[\varepsilon_1(\ti\xi_m)]} \mathrm{Li}_3\left[r(\varepsilon_1,\varepsilon_+)r(\varepsilon_1,\varepsilon_-)\right](\ti\xi_m) \,,
\end{equation}
the force density on a slab \eqref{eq:fd} becomes
\begin{equation}
	f_{\text{A[M]}}(l)=-\frac{H_{\text{A[M]}}}{6\pi l^3} \,.
\label{eq:result}
\end{equation}
We have introduced the Matsubara sum $\sum{\vphantom{\sum}}'$, the Matsubara frequencies $\xi_m=m\,2\pi\kb T/\hbar$, the reflection coefficients $r(\varepsilon_1,\varepsilon_\pm)$, Eq.~\eqref{eq:rp}, the dielectric functions of the right and left slabs $\varepsilon_\pm$, and the dielectric function of the medium between both slabs $\varepsilon_1$. The two results only differ in terms of the additional factor $1/\varepsilon_1$ which arises only in the Maxwell case, written in square brackets. The Hamaker constant for the Abraham theory is in consensus with the definition by Hough and White\cite{HOUGH19803}. The Hamaker constants contain all information about the material properties of the system. In order to compare both methods with experimental data, we need to consider possible errors related to this model. We can identify three potential sources of errors: 

(i) Errors due to imprecise modeling of the dielectric functions of the considered materials. To estimate the uncertainties in the Hamaker constants arising from uncertainties in the dielectric functions, we propagate estimated uncertainties in the parameters of the dielectric functions to uncertainties in the Hamaker constants. Thereby, we make the assumption that the uncertainties in the different parameters of a dielectric function are independent from each other. Further information on the parameter uncertainty for the dielectric functions can be found in section~\ref{sec:modeling_dielectric_functions}. For the Abraham stress tensor, the propagation of uncertainties results in
\begin{eqnarray}
&& 
\left(\frac{\Delta H_{\rm A}}{3\kb T}\right)^2 = \sum_{l} \left(\mats \frac{\mathrm{Li}_2\left[r(\varepsilon_1,\varepsilon_+)r(\varepsilon_1,\varepsilon_-)\right]}{r(\varepsilon_1,\varepsilon_+)r(\varepsilon_1,\varepsilon_-)}  \frac{(\varepsilon_1^2-\varepsilon_-\varepsilon_+)(\varepsilon_-+\varepsilon_+)}{(\varepsilon_++\varepsilon_1)^2(\varepsilon_-+\varepsilon_1)^2} \Delta \tilde\varepsilon_1^{(l)}\right)^2\nonumber\\
&&+  \sum_{n}\left(\mats \frac{\mathrm{Li}_2\left[r(\varepsilon_1,\varepsilon_+)r(\varepsilon_1,\varepsilon_-)\right]}{r(\varepsilon_1,\varepsilon_+)}  \frac{\varepsilon_1}{(\varepsilon_++\varepsilon_1)^2} \Delta \tilde\varepsilon_+^{(n)}\right)^2\nonumber\\
&&+  \sum_{o}\left(\mats \frac{\mathrm{Li}_2\left[r(\varepsilon_1,\varepsilon_+)r(\varepsilon_1,\varepsilon_-)\right]}{r(\varepsilon_1,\varepsilon_-)}  \frac{\varepsilon_1}{(\varepsilon_-+\varepsilon_1)^2} \Delta \tilde\varepsilon_-^{(o)}\right)^2 \, , \label{eq:DeltaHA}
\end{eqnarray}    
where the labels $l$, $n$ and $o$ refer to the different model parameters appearing in the dielectric functions of the medium and the two slabs respectively (see Sec.~\ref{sec:modeling_dielectric_functions}). For the Maxwell case, the result reads
\begin{eqnarray}
\lefteqn{\left(\frac{\Delta H_{\rm M}}{3\kb T}\right)^2 = \sum_{l}\left(\mats \frac{\mathrm{Li}_2\left[r(\varepsilon_1,\varepsilon_+)r(\varepsilon_1,\varepsilon_-)\right]}{\varepsilon_1 r(\varepsilon_1,\varepsilon_+)r(\varepsilon_1,\varepsilon_-)}  \frac{(\varepsilon_1^2-\varepsilon_-\varepsilon_+)(\varepsilon_-+\varepsilon_+)}{(\varepsilon_++\varepsilon_1)^2(\varepsilon_-+\varepsilon_1)^2} \Delta\tilde\varepsilon_1^{(l)} \right. }\nonumber\\
&&\left. -\frac{\mathrm{Li}_3\left[r(\varepsilon_1,\varepsilon_+),r(\varepsilon_1,\varepsilon_-)\right]}{2\varepsilon_1^2 }\Delta\tilde\varepsilon_1^{(l)} \right)^2 +  \sum_{n}\left(\mats \frac{\mathrm{Li}_2\left[r(\varepsilon_1,\varepsilon_+)r(\varepsilon_1,\varepsilon_-)\right]}{\varepsilon_1 r(\varepsilon_1,\varepsilon_+)}  \frac{\varepsilon_1}{(\varepsilon_++\varepsilon_1)^2} \Delta \tilde\varepsilon_+^{(n)}\right)^2\nonumber\\
&&+ \sum_{o} \left(\mats \frac{\mathrm{Li}_2\left[r(\varepsilon_1,\varepsilon_+)r(\varepsilon_1,\varepsilon_-)\right]}{\varepsilon_1 r(\varepsilon_1,\varepsilon_-)}  \frac{\varepsilon_1}{(\varepsilon_-+\varepsilon_1)^2} \Delta \varepsilon_-^{(o)}\right)^2 \, . \label{eq:DeltaHM}
\end{eqnarray}

(ii) The assumption of the nonretarded limit. In the experiments, the force is measured at different separations. The nonretarded Casimir force between the bodies, based on Eq.~(\ref{eq:result}), is then fitted to these measurements and the Hamaker constant is determined from the fit. However, when one considers a too large distance range, the nonretarded Casimir force is not a good approximation anymore, since retardation effects become important. In this case, the determined Hamaker constant might be erroneous.

We analyse this source of error by fitting the nonretarded Casimir force density, Eq.~(\ref{eq:result}), to calculated exact force densities. Those arise from the exact quantum-mechanical expectation values of the two stress tensors, Eqs.~\eqref{eq:result_abraham_full} and \eqref{eq:result_maxwell_full}. We constrain this analysis on the Abraham stress tensor based (standard Lifshitz) theory. 
We estimate a maximum separation between the bodies for each experiment considered, above which the obtained Hamaker constants differ more than 5\,\% from the nonretarded ones. This maximum separation strongly depends on the fitting method that is chosen. The force typically spans several orders of magnitude in experiments (between 1\,nm and 10\,nm, to force density in Eq.~(\ref{eq:result}) already declines by a factor of $10^{-3}$). A direct least-squares based fit using Eq.~(\ref{eq:result}) is thus more controlled by the close-distance measurements that can be expected to have larger quadratic deviations from the fit function. When on the other hand following the approach by~\cite{MILLING1996460}, transforming the force density (Eq.~\ref{eq:result}) to be a linear function of $l$,
\begin{equation}
    \left(\frac{1}{f}\right)^{1/3}=\left(\frac{6\pi}{H}\right)^{1/3}\cdot l\,,
    \label{eq:millingfit}
\end{equation}
measurements at larger distances should have a larger impact on the estimate of the Hamaker constant --- potentially leading to retardation effects altering the estimate. Using this fitting approach, we calculate the maximum separations for the experiments performed by \cite{MILLING1996460} (Table~\ref{tab:maxdist}). The authors of Ref.~\cite{MILLING1996460} expect retardation effects to become important at 5-10\,nm. This holds well for bromobenzene and p-xylene as media (Table~\ref{tab:maxdist} and Fig. 2 in \cite{MILLING1996460}). However, retardation becomes important at 1\,nm distance or less for perfluorohexane and cyclohexane and could thus be an important source of error for these experiments.  

\begin{table}[t]
    \centering
    \setlength\extrarowheight{5pt}
    \begin{tabular}{cc} \hline
         Materials & $d_{\rm max}$  \\ \hline\hline
         Gold, cyclohexane, PTFE&  $<$ 1\,nm  \\
         Gold, p-xylene (benzene), PTFE& 12\,nm  \\
         Gold, bromobenzene, PTFE& 8\,nm  \\
         Gold, perfluorohexane, PTFE& 1\,nm  \\ \hline
    \end{tabular}
    \caption{The maximum distances above which the errors in the fitted Hamaker constants that arise from retardation effects exceed 5\,\% for the materials used in the experiments by \cite{MILLING1996460}. Following to their approach, we use Eq.~(\ref{eq:millingfit}) as fit function.}
    \label{tab:maxdist}
\end{table}

(iii) The application of the proximity force approximation (PFA). The proximity force approximation assumes that the force between objects of arbitrary shape can be approximated on the basis of the force density between two slabs when the objects are close to each other. Mathematically, one then assumes that the force density on the surface of a body is locally given by that for the geometry of slabs, Eq.~(\ref{eq:result}) with the distance defined by the distance to the opposite surface on the second object. One then integrates this force density over the surface of the body. This approximation is used in all experiments considered here to predict the Hamaker constants based on measured forces between spheres and slabs. To assess up to which distance the proximity force approximation holds we consider the simple case of pairwise interactions between molecules in the sphere and molecules in the slab (see \cite{hamaker}), allowing us to explicitly calculate the force for this geometry. We find that proximity force approximation deviates less than 5\,\% up to sphere--slab distances of one percent of the sphere's radius. With sphere radii in the order of micrometers, the proximity force approximation thus should hold up to distances of tens of nanometers. At such distances, the nonretarded approximation is usually not valid anymore as can be seen in table~\ref{tab:maxdist}. It is thus sufficient to verify that the nonretarded approximation is valid in an experimental setup.

\section{Modelling of the dielectric functions}\label{sec:modeling_dielectric_functions}
\begin{figure}[t]
\centering
    \includegraphics[width=0.6\columnwidth]{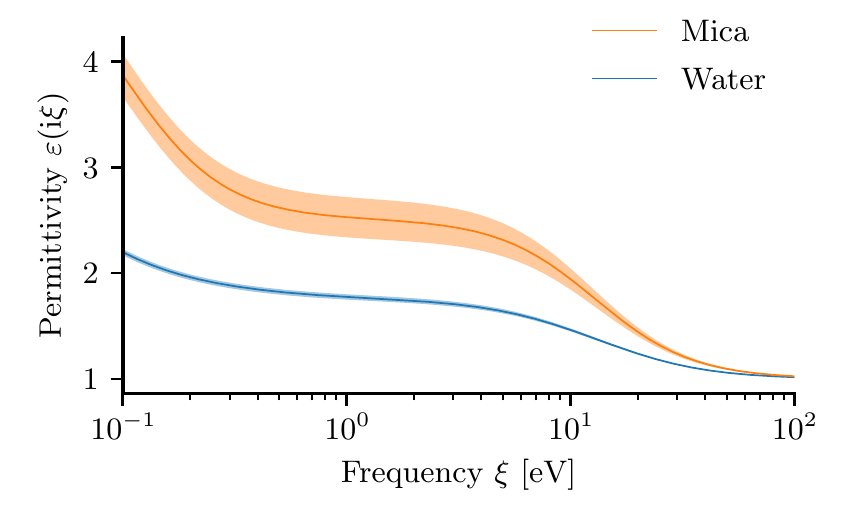}
    \caption{Dielectric functions for mica~\cite{senden} and water~\cite{parsegian} with their associated uncertainty ranges (colored areas) from the relative parameter uncertainties given in table~\ref{tab:relerrors}.}
    \label{figure1}
\end{figure}
Within this study, we compare the predicted Hamaker constants of two theoretical frameworks with the experimental values. These Hamaker constants are calculated on the basis of the dielectric functions of the involved materials. It is thus crucial that these dielectric functions adequately represent the materials. The experimental data are typically fitted to an oscillator model
\begin{equation}
\varepsilon(\ti\xi)=1+\frac{d}{1+ \tau\xi }+ \sum\limits_{k=1}^n\frac{\omega_{p,k}^2}{\omega_{t,k}^2+\xi^2+\xi \gamma_k}\,,\label{eq:osc}
\end{equation}
which is a superposition of Lorentz response functions including a Debye term with strength $d$ and  relaxation time $\tau$. Each of the oscillators has a plasma frequency $\omega_{p,k}$, a resonance frequency $\omega_{t,k}$ and a damping constant $\gamma_k$, which can be damped ($\gamma_k\neq 0$) or undamped ($\gamma_k= 0$). The parameter values for all oscillator models used in this study can be found in Table~\ref{tab:modells}. 

We estimate a relative uncertainty $\delta_p=\Delta p /p$ for the parameters of each material which allows us to estimate the uncertainties of calculated Hamaker constants for both theories. For simplicity, we assume that all parameters of a material have the same independent relative error $\delta$, except for water where we assign a different value to the parameters in the Debye term.  To obtain the relative uncertainties of the parameters, we assess uncertainties in the dielectric functions at different frequencies i$\xi$ and map these to uncertainties in the parameters using propagation of uncertainty  
\begin{equation}
    \left[\Delta\varepsilon(\ti\xi)\right]^2 = \sum_{k=1}^n \left(\Delta\tilde\varepsilon^{(k)}\right)^2 \,,
    \label{eq:uncert1}
\end{equation}
where $n$ denotes the total number of parameters in the dielectric function. The contributions to the uncertainty in the dielectric function from the different parameters depends on the type of parameter. For a plasma frequency it is given by     
\begin{equation}
\Delta\tilde\varepsilon^{(k)} =     \frac{2\omega_{p,k}}{\omega_{t,k}^2+\xi^2+\xi \gamma_k}\Delta \omega_{p,k}\,,
\label{eq:uncert2}
\end{equation}
for a resonance frequency
\begin{equation}
  \Delta\tilde\varepsilon^{(k)} =  \frac{2\omega_{p,k}^2 \omega_{t,k}}{\left(\omega_{t,k}^2+\xi^2+\xi \gamma_k\right)^2}\Delta\omega_{t,k}\,,
  \label{eq:uncert3}
\end{equation}
 for a damping constant
\begin{equation}\Delta\tilde\varepsilon^{(k)} = \frac{\omega_{p,k}^2 \xi}{\left(\omega_{t,k}^2+\xi^2+\xi \gamma_k\right)^2}\Delta\gamma_{k} \, ,
\label{eq:uncert4}
\end{equation}
for a Debye strength
\begin{equation}
\Delta\tilde\varepsilon^{(k)} = \frac{\Delta d}{1+\tau\xi} \, ,
\label{eq:uncert5}
\end{equation}
and for a relaxation time
\begin{equation}
    \Delta\tilde\varepsilon^{(k)} = \frac{d \xi}{\left(1+\tau\xi\right)^2}\Delta\tau \, .
    \label{eq:uncert6}
\end{equation}

For an oscillator model without Debye term, Eqs.~(\ref{eq:uncert1})--(\ref{eq:uncert4}), an uncertainty in the dielectric function at an imaginary frequency $\xi$ can be identified with a relative parameter uncertainty that is given by 
\begin{equation}
 \delta=\frac{\Delta \varepsilon(\ti\xi)}{2\sqrt{\sum_{k}^{}\left[1+\left(\frac{\omega_{t,k}^2}{\omega_{t,k}^2+\gamma_k\xi+\xi^2}\right)^2+\frac{1}{4}\left(\frac{\xi\gamma_k}{\omega_{t,k}^2+\gamma_k\xi+\xi^2}\right)^2\right]\left(\frac{\omega_{p,k}^2}{\omega_{t,k}^2+\gamma_k\xi+\xi^2}\right)^2}}\,.
 \label{eq:uncert7}
\end{equation}
To derive these relative parameter uncertainties $\delta$ for the materials considered in this study, we analyse deviations between the dielectric functions and experimental data as well as deviations between dielectric functions of different material samples. The results can be found in Table~\ref{tab:relerrors}. Uncertainty ranges for the dielectric functions in the mica-water-mica system are depicted in Fig.~\ref{figure1}. The approaches for the materials are listed in the following.

\textbf{Water:} We use the dielectric function from Parsegian~\cite{parsegian}, which is based on a Debye term, five damped infrared oscillators, and six damped UV oscillators. We assign a relative uncertainty for the two Debye parameters based on the deviation of the imaginary part of the dielectric function at real frequencies from absorption measurements performed by \cite{kaatze}. Obtaining a relation similar to Eq.~(\ref{eq:uncert7}) for each real frequency $\omega$, we average the result over all frequencies used by \cite{kaatze}. Likewise, the uncertainty for the parameters in the Lorentz response part, representing the IR and UV spectra, is obtained by analysing the deviation from the UV absorption measurements performed by \cite{hayashi}.

\textbf{Gold:} We use the dielectric function from \cite{parsegian} based on \cite{hagemann}. It uses three damped UV oscillator terms and one term with zero resonance frequency and damping constant which diverges towards zero frequency. Experiments involving gold are performed with bodies coated with gold layers. Depending on the underlying material and the thickness of the gold layer, optical properties vary between experiments. \cite{svetevoy} have fitted Drude-type dielectric functions to samples varying in layer thickness and underlying material, allowing us to assess an uncertainty arising from variation in underlying material and coating thickness. Based on Eq.~\eqref{eq:uncert7}, we calculate an averaged parameter uncertainty $\delta$ arising from half of the spread in the sample Drude functions for Matsubara frequencies up to 3$\cdot10^{15}$\,eV. Up to that frequency, the dielectric function from \cite{parsegian} lies inside the spread given by the Drude functions. We can hence assume that the Drude functions appropriately represent the materials up to this frequency.

\textbf{PTFE:} We use the dielectric function from \cite{ZWOL} which is based on four undamped IR and four undamped UV oscillators. We identify variations in PTFE density between experiments from impurities as the main source of uncertainty for the dielectric function of PTFE. In fact, \cite{MILLING1996460} report the refractive index for PTFE to be $1.32\pm0.02$, which approximately corresponds to a range of PTFE densities between 1.67\,g$/\text{cm}^3$ and 2.1\,g$/\text{cm}^3$~\cite{drummond}. Note that this range also includes the compounds Teflon-AF1600 and Teflon-AF2400 that have very similar dielectric responses compared to PTFE with a first UV absorption at $\approx$\,8\,eV and the second at $\approx$\,20\,eV~\cite{ZWOL}. The Lorentz-Lorenz mixing scheme~\cite{Aspnes} connects the dielectric function $\tilde{\varepsilon}$ at a varied density $\tilde{\rho}$ to a reference dielectric function $\varepsilon$ at a reference density $\rho$ through the relation
\begin{equation}
    \frac{\tilde{\varepsilon}-1}{\tilde{\varepsilon}+2} = \frac{\tilde{\rho}}{\rho} \frac{\varepsilon-1}{\varepsilon +2} \,.
    \label{eq:uncert8}
\end{equation}
Using this relation, one can calculate the uncertainty in the dielectric function of PTFE $\Delta\varepsilon$ resulting from the uncertainty in density
\begin{equation}
    \Delta\varepsilon(\ti\xi)=(\varepsilon(\ti\xi)-1)(\varepsilon(\ti\xi)+2)\frac{\Delta\rho}{3\rho}\,.
    \label{eq:uncert9}
\end{equation}
Using half of the PTFE density range from above as $\Delta\rho$ and inserting \eqref{eq:uncert9} in \eqref{eq:uncert7}, we calculate $\delta$ as the average over all Matsubara frequencies up to a cutoff frequency. This cutoff frequency is determined as the Matsubara frequency above which the Hamaker constants for all experiments with PTFE considered here do not differ more than 10\,\% anymore (for simplicity, we use the Abraham Hamaker constants to determine these cutoff frequencies).

Additionally, one needs to take into account that the dielectric function for PTFE from \cite{ZWOL}, with a refractive index of 1.3433, resembles that of PTFE with density of 2.1\,g/cm$^3$~\cite{drummond}, so at the upper end of the density range derived for the experiment by \cite{MILLING1996460}. Hence, we use Eq.~(\ref{eq:uncert8}) to calculate the Hamaker constants for systems with PTFE based on PTFE corrected to density 1.885\,g/cm$^3$ (the central density in the range 2.1\,g$/\text{cm}^3$ - 1.67\,g$/\text{cm}^3$). Nontheless, the propagated uncertainty for these Hamaker constants is calculated with the uncorrected dielectric function for PTFE, since the uncertainty propagation relies on paramatrized oscillator models of the form (\ref{eq:osc}).

\textbf{Cyclohexane and bromobenzene:} We use the dielectric functions from~\cite{ZWOL} with one undamped IR and four undamped UV oscillators (cyclohexane) and three undamped IR and four undamped UV oscillators (bromobenzene). Due to the large number of oscillators and since they represent a fit to data over a wide range of frequencies, we assign only small uncertainties to the dielectric functions of the two materials. These uncertainties are based on the measured zero-frequency standard deviations given by \cite{ZWOL} and obtained by using Eq.~\eqref{eq:uncert7} for $\xi=\gamma=0$. For bromobenzene, we manually set the dielectric constant to the experimental value of 5.37~\cite{ZWOL} due to the large deviation between the measured values and the modeled dielectric constant. \cite{ZWOL} state that their oscillator models are valid in the range 10$^{-2}$\,eV - 10$^{2}$\,eV which excludes the dielectric constant. However, a significant deviation between the measured dielectric constants and the modeled ones as given by \cite{ZWOL} is only observed for bromobenzene.

\textbf{Mica:} We use the dielectric function from \cite{senden}, containing three undamped infrared oscillators and one undamped UV oscillator. Plasma and resonance frequency for the latter are effective values calculated using the Cauchy plot method and can not be expected to appropriately represent the materials properties over the whole UV range, which is crucial to the calculation of Hamaker constants. With Eq.~\eqref{eq:uncert7}, we calculate the parameter uncertainty for mica as the averaged uncertainty arising from the half spread between the dielectric function from \cite{senden} and the three dielectric functions provided by \cite{parsegian} over Matsubara frequencies up to a cutoff frequency that is again calculated as the frequency above which the Hamaker constant of the mica-water-mica system is not varying more than 10\,\% anymore. The dielectric functions vary with respect to their effective UV resonance frequencies, which lie between 10.33 and 15.66 eV, and thereby account for the uncertainty in the UV.

\textbf{Rutile:} We use the dielectric function that is averaged over the ordinary and extra-ordinary crystallographic axes as given by \cite{buscall,bergstroem}. It contains one IR and one UV oscillator and is based on the Cauchy plot method. Since we do not have full spectral data for rutile, we analyse the spread between different measured and DFT-based values for the zero-frequency permittivity as well as for the real part of the dielectric function in the visible\cite{dou,parker,davis}. The uncertainty in the visible is larger and is used for the relative parameter uncertainty of rutile in this study.

\textbf{P-xylene:} A dielectric function for p-xylene is provided by \cite{masuda}. However, it is again based on the Cauchy plot method with only one UV oscillator, and hence it does not appropriately represent the properties in the UV.  We instead make use of the fact that p-xylene is chemically similar both to benzene and toluene. We use benzene from \cite{ZWOL} as the dielectric function for p-xylene\footnote{Analogously, one could also choose toluene from \cite{ZWOL} to represent the dielectric function of p-xylene. This choice only insignificantly changes the resulting Hamaker constants with respect to those obtained using benzene. This is also reflected by the small parameter uncertainty we assign to p-xylene (Table~\ref{tab:relerrors}).}. The parameter uncertainty is obtained as the averaged uncertainty arising from Eq.~\eqref{eq:uncert7}, assuming the uncertainty in the dielectric function to be given by the deviation from the dielectric function for toluene~\cite{ZWOL}, over Matsubara frequencies up to a cutoff frequency that is defined analogously as before.   

\textbf{Perfluorohexane:} A dielectric function for perfluorohexane was published by \cite{drummond}. As before, their dielectric function only has one UV oscillator. As argued by~\cite{ZWOL} the oscillator frequency is chosen such that the permittivity cuts off too early in the UV. Despite this shortcoming, we use it in our study since it is the only published dielectric function for perfluorohexane to our knowledge. To estimate the parameter uncertainty, we create an alternative dielectric function for perfluorohexane based on that for PTFE \cite{ZWOL} by rescaling the density using Eq.~\ref{eq:uncert8} to the perfluorohexane density $\rho=1.6995$\,g/cm$^3$ that is reported by \cite{drummond}. Following the methodology by \cite{ZWOL}, we rescale the dielectric function obtained this way such that it fits the dielectric constant reported by \cite{drummond}. The derived dielectric function for perfluorohexane has a similar refractive index to that reported by \cite{drummond}, differing on the third decimal place. The parameter uncertainty is then finally obtained using Eq.~\eqref{eq:uncert7} where the uncertainty in the dielectric function is identified with the deviation between the dielectric function by \cite{drummond} and that derived here. The large deviation between the two leads to the largest parameter uncertainty estimated here equal to 15.1\,\%. As before, it is calculated as the averaged uncertainty over the Matsubara frequencies up to a cutoff frequency above which the Hamaker constant of the system with perfluorohexane is not varying more than 10\,\% anymore. 

\begin{table}[t!]
    \centering
    \setlength\extrarowheight{7.5pt}
    \begin{tabular}{ccc} \hline
        Material & Rel. error $\delta$  & Method \\ \hline \hline
        Water-Debye & 9.3\,\% & \parbox[c]{8cm}{Model deviation from microwave data~\cite{kaatze}} \\[7.5pt] \hline 
        Water-IR/UV & 2.5\,\% & \parbox[c]{8cm}{Model deviation from UV data~\cite{hayashi}} \\[7.5pt] \hline
        Gold & 12.8\,\% & \parbox[c]{8cm}{Spread between gold film samples over IR frequencies~\cite{svetevoy}} \\[7.5pt] \hline
        PTFE & 8.4 \,\% & \parbox[c]{8cm}{Uncertainty from different PTFE densities over IR and UV frequencies~\cite{drummond}} \\[7.5pt] \hline
        Cyclohexane & 1.1 \,\% & \parbox[c]{8cm}{Zero-frequency uncertainty of permittivity~\cite{ZWOL}} \\[7.5pt] \hline
        Bromobenzene & 1.6 \,\% & \parbox[c]{8cm}{Zero-frequency uncertainty of permittivity~\cite{ZWOL}} \\[7.5pt] \hline
        Mica & 4.5 \,\% & \parbox[c]{8cm}{Spread between dielectric functions over IR and UV frequencies~\cite{parsegian,senden}} \\[7.5pt] \hline
        Rutile & 4.2 \,\% & \parbox[c]{8cm}{Spread in DFT permittivities in the visible~\cite{dou,parker,davis}} \\[7.5pt] \hline
        p-Xylene & 1.6 \,\% & \parbox[c]{8cm}{Deviation between benzene and toluene permittivity over IR and UV~\cite{ZWOL}} \\[7.5pt] \hline
        Perfluorohexane & 15.1 \,\% & \parbox[c]{8cm}{Deviation from density-rescaled PTFE permittivity over IR and UV frequencies~\cite{ZWOL}} \\[7.5pt] \hline
    \end{tabular}
    \caption{Relative errors on the oscillator parameters of the materials.}
    \label{tab:relerrors}
\end{table}

\section{Results and Discussion}
\label{Sec4}
In order to compare with the theoretical predictions, we chose all experimental studies to our knowledge that quoted experimental uncertainties on the Hamaker constants\cite{JG78,MILLING1996460,larson1993,walsh2012,craig_pers_com}.  We investigate whether, given the uncertainties within the predicted as well as the measured Hamaker constants, it is possible to distinguish between the two theories based on experiments.  

The main focus of this work is to thoroughly assess the uncertainties within the theoretical predictions and to assign uncertainty ranges to the predicted Hamaker constants. To this end, we presented a detailed formalism to obtain uncertainties for theoretical predictions in Secs.~\ref{section:theory} and \ref{sec:modeling_dielectric_functions}. Errors in the dielectric functions involved, arising from error-prone representation of the materials and insufficient knowledge about the samples used in the experiments, represent the main source of uncertainty and are quantified using a two-step procedure. First, we assign relative uncertainties to the parameters within the dielectric functions. Using the propagation of uncertainty framework, these are then mapped to uncertainties in the predicted Hamaker constants.

	\begin{table}[h!]
	\setlength\extrarowheight{7.5pt}
		\centering
		\begin{tabular}{ccccccc} \hline
			{\small\textbf{Materials}} & \multicolumn{2}{c}{\small\textbf{Abraham}} &\multicolumn{2}{c}{\small\textbf{Maxwell}}  &
			\multicolumn{2}{c}{\small\textbf{Experiment}}\\
			& {\small$H_{\rm A}$}& {\small$\Delta H_{\rm A}$[\%]} & {\small$H_{\rm M}$}& {\small$\Delta H_{\rm M}$[\%]} & {\small$H_{\rm E}$}& {\small$\Delta H_{\rm E}$[\%]} \\ \hline\hline
			{\small Mica - water - mica} 		& {\small 2.00 $\pm$ 0.47} 	& {\small 24} &{\small  1.17 $\pm$ 0.31} 	& {\small 27}  & {\small 2.2 $\pm$ 0.3 \cite{JG78}}& {\small 14}\\
			{\small Gold - cyclohexane - PTFE} 	& {\small -1.57 $\pm$ 1.11}	& {\small 71} &{\small  -0.73 $\pm$ 0.70}  	& {\small 96} &{\small  -5.5 $\pm$ 1.5 \cite{MILLING1996460}} &{\small 27}\\
            {\small Gold -  benzene - PTFE}	 		& {\small -2.88 $\pm$ 1.06} 	& {\small 37}	& {\small -1.40 $\pm$ 0.62}	& {\small 44} & {\small -5 $\pm$ 1 \cite{MILLING1996460}} & {\small 20} \\
			{\small Gold - bromobenzene - PTFE}	& {\small -3.98 $\pm$ 1.01}	& {\small 25}	& {\small -1.86 $\pm$0.57} 	& {\small 31} & {\small -5.5 $\pm$ 1.5 \cite{MILLING1996460}} & {\small 27}\\
			{\small Gold - perfluorohexane - PTFE}	& {\small 1.95 $\pm$ 3.14}	& {\small 161} & {\small 1.55  $\pm$ 2.09}	& {\small 134} & {\small 0.7 $\pm$ 0.3 \cite{MILLING1996460}} & {\small 43}\\
			{\small Rutile - water - rutile} 	& {\small 6.21 $\pm$ 0.65} 	& {\small 10} & {\small 3.64 $\pm$ 0.38} 	& {\small 10} & {\small 6 $\pm$ 2 \cite{larson1993}} & {\small 33} \\
			&  	&  &   	&  & {\small 6 $\pm$ 1 \cite{walsh2012,craig_pers_com}} & {\small 17} \\\hline
		\end{tabular}                                                                 
		\caption{Comparison of the Hamaker constants obtained for the two theories to experimental values (in $10^{-20}\,\rm{J}$). Uncertainty ranges are given for the theoretical Hamaker constants as derived in this study and experimental uncertainty ranges are taken from the experimental studies. The experimental studies are referenced. The second experimental value  for the rutile-water-rutile system is from \cite{walsh2012}. The uncertainty on their experimental Hamaker constant is obtained from private communication~\cite{craig_pers_com}.}                              
		\label{tab:results}
	\end{table} 

Casimir forces in media do not only pose additional experimental challenges but also require more detailed knowledge of the dielectric functions of the materials for theoretical calculations. Consequently, we generally estimate large uncertainty ranges for the theoretically predicted Hamaker constants that result from the uncertainties we assigned to the involved dielectric functions. Those arise either from insufficient representation of the materials properties as in the case of perfluorohexane, p-xylene, mica, and rutile or from large variations between samples of the same material as in the case of gold and PTFE. Insufficient representation is often a consequence of determining the dielectric function mainly from refractive index data using the Cauchy plot method which tends to not appropriately represent the properties in the UV spectrum. Sample variations for PTFE are due to varying densities between PTFE samples and gold samples usually depend on coating thickness and the underlying substrate. In these cases, one should ideally determine the dielectric functions from the samples also used in the experiments to reduce uncertainties when comparing to theory. Alternatively, experimental studies should provide the information necessary to make theoretical predictions for the specific samples, such as density or high-precision refractive index measurements for PTFE. Repulsive Hamaker constants often involve two dielectric functions that are close to each other and tend to be more sensitive to uncertainties in the dielectric functions, in particular when the Hamaker constants are close to zero.

\begin{figure}[t!]
    \centering
    \includegraphics[width=0.7\columnwidth]{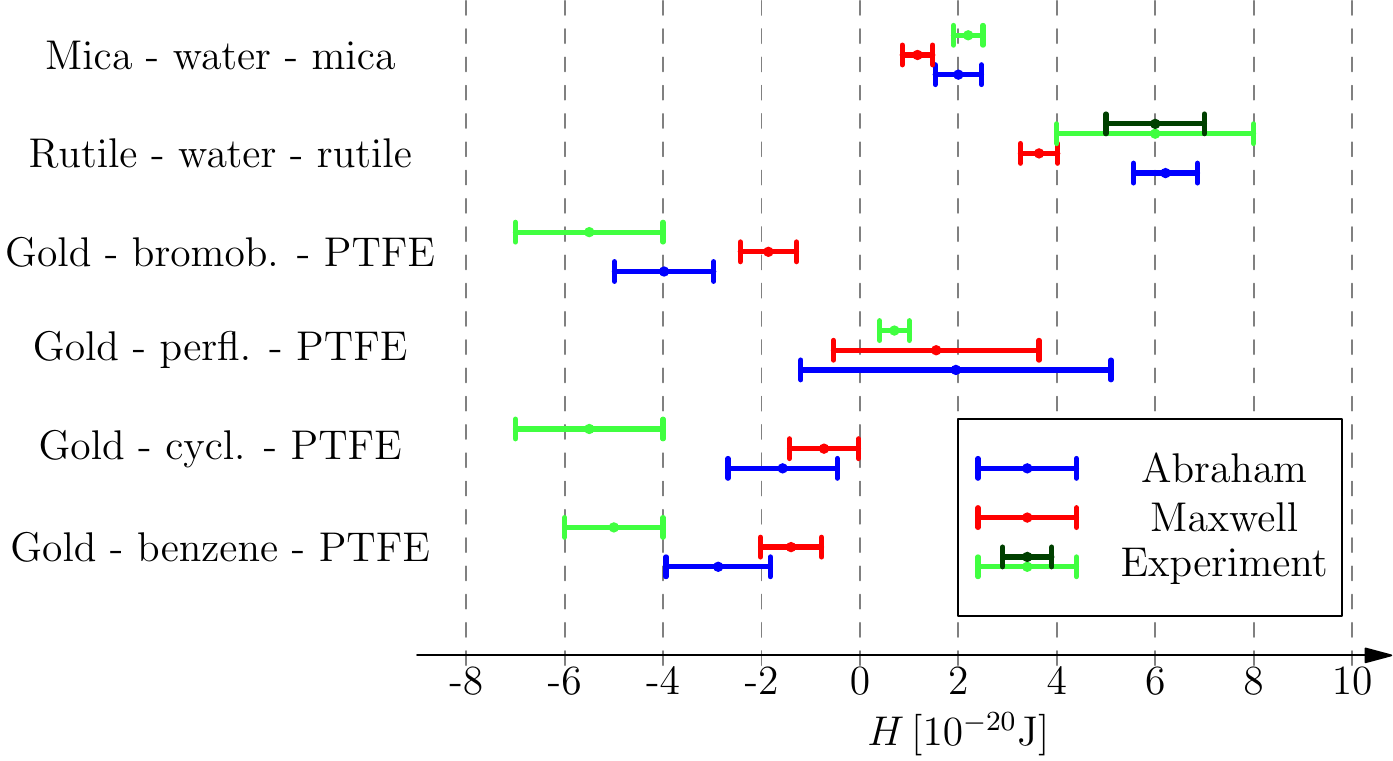}
    \caption{Experimentally determined Hamaker constants in comparison to the ones obtained from the theories based on the Abraham and the Maxwell stress tensors. The dots represent the measured/calculated values while the bars represent the uncertainty ranges.}
    \label{figure2}
\end{figure}

We now compare the theoretical predictions that arise from the two Casimir force theories to experimental results. The results are given in Table~\ref{tab:results} and depicted in Fig.~\ref{figure2}. We generally see a tendency that predictions based on the Abraham stress tensor (Lifshitz theory) are closer to the experimental values than the ones based on the Maxwell stress tensor. Standard Lifshitz theory agrees, within uncertainty ranges, with the mica-water-mica, gold-bromobenzene-PTFE, gold-perfluorohexane-PTFE, and both rutile-water-rutile experiments~\cite{JG78,larson1993,MILLING1996460,walsh2012,craig_pers_com}. The Maxwell stress tensor based theory on the other hand only agrees with the gold-perfluorohexane-PTFE and one of the two rutile-water-rutile experiments~\cite{MILLING1996460,larson1993}. We view the mica-water-mica experiment as a strong evidence for the Lifshitz theory in our study, since this experiment was performed with a surface force apparatus that is viewed to be more precise than atomic force microscopes. As an aside, note that the difference between the two predictions is in this case considerably influenced by the zero-frequency Matsubara term: it contributes 13.3\% to the overall Hamaker constant in the Abraham approach while its contribution is strongly suppressed to only 0.3\% in the Maxwell approach due to the high static dielectric constant of water. The experiment with perfluorohexane as medium is not suitable to decide between the two theories due to the large uncertainty we assign to the calculated Hamaker constants based on the poor representation of the dielectric function of perfluorohexane.

The two repulsive gold-PTFE experiments with cyclohexane and p-xylene as media tend to have more negative Hamaker constants than we predict with both theories. While the uncertainty ranges for the p-xylene system almost overlap for the Lifshitz theory, there is a large deviation for the cyclohexane system. The Hamaker constants for all systems with PTFE considered here are very sensitive to variations in PTFE density. A possible part of the explanation would be that the density of the PTFE sample used by \cite{MILLING1996460} was on the lower end of the estimated density range, so at 1.67\,g/cm$^3$. When calculating the Hamaker constants for these systems with a dielectric function for PTFE that is adjusted to 1.67\,g/cm$^3$ (using Eq.~\ref{eq:uncert8}), Lifshitz theory additionally also agrees with the p-xylene experiment (the theoretical uncertainty range, $-3.84\pm0.21\cdot10^{-20}$\,J, now only derived from the uncertainties in p-xylene and gold representations, overlaps with the experimental range). Even lower PTFE densities would be required to match the experimental value for the system with cyclohexane. However, as we have seen in Sec.~\ref{section:theory}\,ii), experimentally determined Hamaker constants for cyclohexane are likely to have a retardation bias that needs to be considered already for distances below 1\,nm. Such a retardation bias should lead to a too negative Hamaker constant and could thus help to explain the remaining missmatch. For the Maxwell case on the other hand, it is not possible to match the experimental results with reasonable values for the density of PTFE.

\section{Conclusions}
In summary, we have derived two alternative expressions for the Hamaker constants from the two contradicting Abraham and Maxwell stress-tensor approaches. We have also quantified all discussed possible errors, which are dominated by uncertainties in the optical constants of the materials involved. The derived models have been compared to a specific set of existing experimental results where a liquid separates both objects and the experimental uncertainties have been considered and reported.
We find close agreement between experiment and Lifshitz theory for the mica-water-mica and rutile-water-rutile systems. In contrast, no agreement can be found with the Maxwell stress tensor based theory for these systems. The three investigated repulsive systems also clearly point towards standard Lifshitz theory, although agreement within the assigned uncertainty ranges is only found for the gold-bromobenzene-PTFE system. Finally, we cannot draw conclusions from the perfluorohexane experiment. Uncertainties in the theoretical predictions are clearly too large in this case, again highlighting the necessity of high-quality and sample-specific dielectric functions to make precise predictions for Casimir force experiments.

\section*{Acknowledgement}

The authors thank F. Intravaia, I. Brevik and M. Bostr\"om for fruitful discussions. We gratefully acknowledge support from the German Research Council (grants BU 1803/6-1, S.Y.B. and J.F., BU 1803/3-1, S.Y.B.) and from the Research Council of Norway (Project  250346, J.F.).

\bibliographystyle{unsrt}  
\bibliography{bib.bib}  %%% Remove comment to use the external .bib file (using bibtex).

%\begin{thebibliography}{1}

%\bibitem{kour2014real}
%George Kour and Raid Saabne.
%\newblock Real-time segmentation of on-line handwritten arabic script.
%\newblock In {\em Frontiers in Handwriting Recognition (ICFHR), 2014 14th
%  International Conference on}, pages 417--422. IEEE, 2014.

%\bibitem{kour2014fast}
%George Kour and Raid Saabne.
%\newblock Fast classification of handwritten on-line arabic characters.
%\newblock In {\em Soft Computing and Pattern Recognition (SoCPaR), 2014 6th
%  International Conference of}, pages 312--318. IEEE, 2014.

%\bibitem{hadash2018estimate}
%Guy Hadash, Einat Kermany, Boaz Carmeli, Ofer Lavi, George Kour, and Alon
%  Jacovi.
%\newblock Estimate and replace: A novel approach to integrating deep neural
%  networks with existing applications.
%\newblock {\em arXiv preprint arXiv:1804.09028}, 2018.

%\end{thebibliography}

\section*{Appendix}
\appendix

\section{Calculation of the quantum-mechanical expectation values of stress tensors and nonretarded limit}\label{app0}
Based on the expressions (\ref{eq:2_1})--(\ref{eq:2_4}), the Casimir force acting on a dielectric body of volume $V$ can be obtained by integrating the stress tensor over the surface of the body~\cite{buh1}
\begin{equation}
\mathbf{F}(t)=\intd{\partial V}{}{\mathbf{A}}\cdot\ev{\mathbf{T}(\mathbf{r},t)}\,,
\label{eq:CF}
\end{equation}
with the quantum mechanical expectation value of the stress tensor $\ev{\mathbf{T}(\mathbf{r},t)}$. We apply the quantization scheme~\cite{buh1} to obtain the quantum-mechanical expressions of the stress tensors for the quantized thermal electromagnetic field. The system's geometry is described by the Green's tensor. Finally, we then choose the geometry to be that of two infinite slabs with an intervening medium and calculate the force density acting on the slabs.

Employing the quantisation scheme introduced in Ref.~\cite{buh1}, one finds for the thermal expectation values of the electric-field stress tensor components
\begin{eqnarray}
\ev{\hat{\mathbf{E}}(\mathbf{r})\hat{\mathbf{E}}(\mathbf{r})}=\frac{\hbar}{\varepsilon_0\pi}\intd{0}{\infty}{\omega}\left[1+2n(\omega)\right]\frac{\omega^2}{\tc^2}\im\mathbf{G}(\mathbf{r},\mathbf{r},\omega)\,, \\
\ev{\mathbf{D}(\mathbf{r})\mathbf{E}(\mathbf{r})}=\frac{\hbar}{\pi}\intd{0}{\infty}{\omega}\left[1+2n(\omega)\right]\frac{\omega^2}{\tc^2}\im\biggl\{\varepsilon(\mathbf{r},\omega)\mathbf{G}(\mathbf{r},\mathbf{r},\omega)\biggr\}\,,
\end{eqnarray}
where $n(\omega)=1/[\exp\left(\hbar\omega/\kb T\right)-1]$ is the thermal photon number. Similar results are found for the magnetic fields. The electromagnetic Green's tensor $\mathbf{G}(\mathbf{r},\mathbf{r}',\omega)$ describes the propagation of the electromagnetic field from a source point $\mathbf{r}'$ to a field point $\mathbf{r}$. For purely dielectric media as considered in this study, it is the solution to the vector Helmholtz equation
\begin{equation}
\left[\nabla\times\nabla\times-\frac{\omega^2}{\tc^2}\varepsilon(\mathbf{r},\omega)\right]\mathbf{G}(\mathbf{r},\mathbf{r}',\omega)=\boldsymbol{\delta}(\mathbf{r}-\mathbf{r}')\,.
\end{equation}
We subtract the bulk part $\mathbf{G}\bu$ of the Green's tensor, which is connected to the position-independent Lamb shift~\cite{lamb} and does not contribute to forces between bodies, and only use the scattering part $\mathbf{G}\scae$ in the following: $\mathbf{G}\to\mathbf{G}-\mathbf{G}\bu=\mathbf{G}\scae$. Performing a rotation on the complex frequency plane, extending the integral to a closed contour and then identifying the only non-vanishing contribution to be the residues connected to the enclosed poles, one finds the mean values of the two competing stress tensors with respect to the thermal electromagnetic field in a homogeneous medium to be given by
\begin{align}
\begin{split}
&\ev{\mathbf{T}\sub{A}(\mathbf{r})}=-2\kb T\mats\Bigg\{\varepsilon\frac{\xi_m^2}{\tc^2}\sca(\mathbf{r},\mathbf{r})+\scacurlt{\mathbf{r}}{\mathbf{r}'}\\
&-\frac{1}{2}\,\mathbb{I}\,\tr\left[\varepsilon\frac{\xi_m^2}{\tc^2}\sca(\mathbf{r},\mathbf{r})+\scacurlt{\mathbf{r}}{\mathbf{r}'}\right]\!\!\Bigg\}\,,\label{eq:TA}
\end{split}
\end{align}
\begin{align}
\begin{split}
&\ev{\mathbf{T}\sub{M}(\mathbf{r})}=-2\kb T\mats\Bigg\{\frac{\xi_m^2}{\tc^2}\sca(\mathbf{r},\mathbf{r})+\scacurlt{\mathbf{r}}{\mathbf{r}'}\\
&-\frac{1}{2}\,\mathbb{I}\,\tr\left[\frac{\xi_m^2}{\tc^2}\sca(\mathbf{r},\mathbf{r})+\scacurlt{\mathbf{r}}{\mathbf{r}'}\right]\!\Bigg\}\,,\label{eq:TM}
\end{split}
\end{align}
where $\xi_m=m\,2\pi\kb T/\hbar$ denotes the Matsubara frequencies and the primed sum, $\sum{\vphantom{\sum}}'$, means that the $0$th term carries half-weight. We see that the two results only differ in an additional factor of $\varepsilon(\ti\xi_m)$, denoting the response function of the homogeneous intervening medium. 

These general theories are now applied to calculate the force between two infinite dielectric slabs with an intervening dielectric medium. This will then allow us to describe forces between close but otherwise arbitrary dielectric objects in media.

For calculating the force on a slab based on Eqs.~(\ref{eq:TA}) and (\ref{eq:TM}), one has to determine the scattering Green's tensor $\sca$ for the geometry of two slabs with an intervening medium, in particular for the case where both position arguments are inside the intervening medium.	We denote the electric response function of the left slab by $\varepsilon_-$, the one of the right slab by $\varepsilon_+$ and the one of the medium by $\varepsilon_1$. The distance between the plates is denoted by $l$. The Green's tensor is given by~\cite{buh1}
\begin{align}
\begin{split}
&\sca(\mathbf{r},\mathbf{r}',\ti\xi)=\frac{1}{8\pi^2}\intdz{}{}{\mathbf{k}^\parallel}
\frac{e^{\ti\mathbf{k}^\parallel(\mathbf{r}-\mathbf{r}')}}{\kappa^\perp}\sum\limits_{\sigma=\rm{s,p}} \Big\{\frac{r_\sigma^+r_\sigma^-e^{-2\kappa^\perp l}}{D_\sigma}\left[\fe_{\sigma}^+\fe_{\sigma}^+e^{-\kapp(z-z')}+\fe_{\sigma}^-\fe_{\sigma}^-e^{\kapp(z-z')}\right]\\
&+\frac{1}{D_\sigma}\!\!\left[\fe_{\sigma}^+\fe_{\sigma}^-r_\sigma^-e^{-\kapp(z+z')}+\fe_{\sigma}^-\fe_{\sigma}^+r_\sigma^+e^{-\kapp(2l-z-z')}\right]\!\!\Big\}\,.
\end{split}
\label{eq:GS_full}
\end{align}
The coefficients p and s denote s- and p-polarized waves  and $\fe_\sigma^+$ and $\fe_\sigma^-$ represent the corresponding polarisation unit vectors. $\mathbf{k}^\parallel$ denotes the wave vector parallel to the surface of the slabs and
\begin{equation}
\kapp\equiv\sqrt{\varepsilon_1\frac{\xi^2}{\tc^2}+\kpars} \, ,
\label{eq:kpara}
\end{equation}
is the imaginary part of the perpendicular projection of the wave vector with parallel part $k^\parallel$. We further introduced the auxiliary quantity $D_p\equiv1-r_p^+r_p^-e^{-2\kapp l}$, which includes the multiple reflections of the electromagnetic waves between both interfaces. The Fresnel reflection coefficients read 
\begin{equation}
r_s^\pm=\frac{\kapp-\kapp_\pm}{\kapp+\kapp_\pm}\,,\quad r_p^\pm=\frac{\kapp\varepsilon_{\pm}-\kapp_\pm\varepsilon_1}{\kapp\varepsilon_\pm+\kapp_\pm\varepsilon_1}\,,
\label{eq:rp_full}
\end{equation} 
with $\kapp_\pm$ being the analogue of Eq.~\eqref{eq:kpara} inside the two slabs, reading
\begin{equation}
	\kapp_\pm=\sqrt{\varepsilon_\pm\frac{\xi^2}{\tc^2}+{k^\parallel}^2}\,.
\end{equation}
Inserting the scattering Green's tensor, Eq.~\eqref{eq:GS_full}, into Eqs.~\eqref{eq:TA} and \eqref{eq:TM}, we obtain the quantum-mechanical expectation values of the stress tensors within the intervening medium as
\begin{equation}
\ev{\mathbf{T}\sub{A}{\vphantom{\mathbf{T}\sub{A}}}_{zz}}=\frac{\kb T}{\pi}\mats\intd{0}{\infty}{k^\parallel}k^\parallel\kapp\sum\limits_{\sigma=s,p}\frac{r_\sigma^+ r_\sigma^- e^{-2\kapp l}}{D_\sigma}\,,
\label{eq:result_abraham_full}
\end{equation}
and
\begin{align}
\begin{split}
\ev{\mathbf{T}\sub{M}{\vphantom{\mathbf{T}\sub{M}}}_{zz}}(z)&=\frac{\kb T}{\pi}\mats\intd{0}{\infty}{k^\parallel}k^\parallel\Big\{\kapp\sum\limits_{\sigma=s,p}\left(\delta_{\sigma s}+\frac{1}{\varepsilon_1}\delta_{\sigma p}\right)\frac{r_\sigma^+r_\sigma^-e^{-2\kapp l}}{D_\sigma}\\
&-\frac{\varepsilon_1\,\xi_m^2}{2\kapp\tc^2}\left(1-\frac{1}{\varepsilon_1}\right)\sum\limits_{\sigma=s,p}\left(\delta_{\sigma s}-\delta_{\sigma p}\right)\frac{2r_\sigma^+r_\sigma^-e^{-2\kapp l}\!+\!r_\sigma^+e^{-2\kapp (l-z)}+r_\sigma^-e^{-2\kapp z}}{2D_\sigma}\!\Big\}\,,
\end{split}
\label{eq:result_maxwell_full}
\end{align}
These results have been found earlier by Brevik and Ellingsen~\cite{brevik_comment}. The expectation value of the Maxwell stress tensor, a function of $z$, depends on the location within the medium. It diverges for $z=0,l$.

We now derive the nonretarded limit of these results, Eqs.~\eqref{eq:result_abraham_full} and \eqref{eq:result_maxwell_full}, that is valid for small separations between the plates compared to the system's relevant transition wavelengths. In the nonretarded limit, one can assume $\frac{\xi l}{\tc},\,\sqrt{\varepsilon_\pm-\varepsilon_1}\frac{\xi l}{\tc}\ll 1$ and $\frac{\xi}{\tc\kapp},\,\sqrt{\varepsilon_\pm-\varepsilon_1}\frac{\xi}{\tc\kapp}\ll 1$. The Fresnel reflection coefficients~\eqref{eq:rp_full} become
\begin{equation}
r_s^\pm\simeq0\,\text{ and }\, r_p^\pm\simeq\frac{\varepsilon_{\pm}-\varepsilon_1}{\varepsilon_\pm+\varepsilon_1} \coloneqq r(\varepsilon_1,\varepsilon_\pm)\,,\label{eq:rp}
\end{equation}
since 
\begin{equation}
	\kapp_\pm=\kapp\sqrt{(\varepsilon_\pm-\varepsilon_1)\frac{\xi^2}{\tc^2\kapps}+1}\simeq\kapp\,,
\end{equation} 
in the nonretarded limit.

Using these assumptions, we obtain the nonretarded approximations of Eqs.~\eqref{eq:result_abraham_full} and \eqref{eq:result_maxwell_full} as
\begin{equation}
\ev{\mathbf{T}\sub{A}{\vphantom{\mathbf{T}\sub{A}}}_{zz}}=\frac{\kb T}{\pi}\mats\intd{0}{\infty}{k^\parallel}k^\parallel\kapp\frac{r_p^+ r_p^- e^{-2\kapp l}}{D_p}\,,
\label{eq:result_abraham}
\end{equation}
and 
\begin{align}
\begin{split}
\ev{\mathbf{T}\sub{M}{\vphantom{\mathbf{T}\sub{M}}}_{zz}}=&\frac{\kb T}{\pi}\mats\intd{0}{\infty}{k^\parallel}k^\parallel \kapp\frac{r_p^+r_p^-e^{-2\kapp l}}{\varepsilon_1 D_p}\,.
\end{split}
\label{eq:result_maxwell}
\end{align}
The force on a slab is finally calculated according to Eq.~\eqref{eq:CF}. As the slabs are infinitely thick, only the surface integral at the interface to the medium contributes and the force density per surface area can be calculated according to Eq.~\eqref{eq:fd}. To obtain Eq.~\eqref{eq:result}, we also substitute the integration variable in Eqs.~\eqref{eq:result_abraham} and \eqref{eq:result_abraham} by $\kappa^\perp$ and realise
that the lower integral bound, $\sqrt{\varepsilon_1}\xi/c$, can be set to zero in the nonretarded limit~\cite{buh1}.
\newpage
\section{Oscillator parameters of the permittivity models}\label{app2}
\begin{table}[h!]
    \setlength\extrarowheight{0pt}
    \begin{minipage}[t]{0.49\textwidth}
    \vspace{0pt}
    \centering
    \begin{tabular}{ccc}\hline
         $\omega_p$\,[eV] \hspace{0.5cm} & $\omega_t$\,[eV] \hspace{0.5cm}  & $\gamma$ [eV] \hspace{0.5cm} \\ \hline
         \multicolumn{3}{c}{\textbf{Water}~\cite{parsegian}}\\ \hline
         \multicolumn{3}{c}{(Debye term: d=74.8, $\tau$=1.5267$\cdot10^{4}$\,eV)}\\ 
         $2.5000\cdot10^{-2}$ & $2.07\cdot10^{-2}$ & $1.5\cdot10^{-2}$ \\
         $5.9161\cdot10^{-2}$ & $6.9\cdot10^{-2}$ & $3.8\cdot10^{-2}$ \\
         $3.5777\cdot10^{-2}$ & $9.2\cdot10^{-2}$ & $2.8\cdot10^{-2}$ \\
         $2.3324\cdot10^{-2}$ & $2\cdot10^{-1}$ & $2.5\cdot10^{-2}$ \\
         $1.1619\cdot10^{-1}$ & $4.2\cdot10^{-1}$ & $5.6\cdot10^{-2}$ \\
         1.6371 & 8.25 & $5.1\cdot10^{-1}$ \\
         2.3812 & 10.0 & $8.8\cdot10^{-1}$ \\
         3.4641 & 11.4 & 1.54 \\
         5.1284 & 13.0 & 2.05 \\ 
         5.8138 & 14.9 & 2.96 \\
         9.6333 & 18.5 & 6.26 \\ \hline
         \multicolumn{3}{c}{\textbf{Gold}~\cite{parsegianweiss}}\\ \hline
         - & 6.3332 & - \\
         3.87 & 7.7208 & 2.62 \\
         8.37 & 11.070 & 6.41 \\
         23.46 & 32.112 & 27.57 \\ \hline
         \multicolumn{3}{c}{\textbf{Rutile}~\cite{bergstroem}}\\ \hline
         4.7883$\cdot10^{-1}$ & 4.6075$\cdot10^{-2}$ & - \\
         10.893 & 4.8379 & - \\ \hline
         \multicolumn{3}{c}{\textbf{PTFE}~\cite{ZWOL}}\\ \hline
         2.8931$\cdot10^{-5}$ & 3$\cdot10^{-4}$ & - \\
         1.0281$\cdot10^{-3}$ & 7.6$\cdot10^{-3}$ & - \\
         2.0766$\cdot10^{-2}$ & 5.57$\cdot10^{-2}$ & - \\
         4.2168$\cdot10^{-2}$ & 1.26$\cdot10^{-1}$ & - \\
         2.9631 & 6.71 & - \\
         12.310 & 18.6 & - \\
         13.707 & 42.1 & - \\
         15.246 & 77.6 & - \\ \hline
    \end{tabular}
    \end{minipage}
    \begin{minipage}[t]{0.49\textwidth}
    \vspace{0pt}
    \centering
    \begin{tabular}{ccc}\hline
         $\omega_p$\,[eV] \hspace{1.25cm}& $\omega_t$\,[eV] \hspace{1.25cm} & $\gamma$ [eV]\hspace{1.25cm} \\ \hline
    \multicolumn{3}{c}{\textbf{Mica}~\cite{senden}}\\ \hline
         6.6217$\cdot10^{-2}$ & 6.4900$\cdot10^{-2}$ & -\\
         6.8913$\cdot10^{-2}$ & 1.0262$\cdot10^{-1}$ & -\\
         9.7498$\cdot10^{-2}$ & 1.1538$\cdot10^{-1}$ & -\\
         1.0173$\cdot10^{-1}$ & 1.2282$\cdot10^{-1}$ & -\\
         15.867 & 12.921 & - \\ \hline  
         \multicolumn{3}{c}{\textbf{Cyclohexane}~\cite{ZWOL}}\\ \hline
         2.6543$\cdot10^{-2}$ & 2.16$\cdot10^{-1}$ & -\\
         2.9395 & 8.03 & -\\
         7.8374 & 10.9 & -\\
         10.162 & 18.4 & -\\
         8.9059 & 41.3 & -\\ \hline
         \multicolumn{3}{c}{\textbf{Bromobenzene}~\cite{ZWOL}}\\ \hline
         1.1709$\cdot10^{-3}$ & 5.02$\cdot10^{-3}$ & - \\
         4.1915$\cdot10^{-3}$ & 3.09$\cdot10^{-2}$ & - \\
         2.4192$\cdot10^{-2}$ & 1.11$\cdot10^{-1}$ & - \\
         4.9233 & 6.75 & - \\
         10.681 & 13.3 & - \\
         11.758 & 24.0 & - \\
         9.6185 & 99.9 & - \\ \hline
         \multicolumn{3}{c}{\textbf{Benzene (for p-Xylene)}~\cite{ZWOL}}\\ \hline
         1.3343$\cdot10^{-1}$ & 8.76$\cdot10^{-1}$ & - \\
         5.6100$\cdot10^{-1}$ & 6.71 & - \\ 
         1.3069 & 4.48 & - \\ 
         8.6182 & 17.0 & - \\
         6.8840 & 8.48 & - \\
         11.077 & 23.3 & - \\ 
         8.3827 & 70.1 & - \\ \hline
         \multicolumn{3}{c}{\textbf{Perfluorohexane}~\cite{drummond}}\\ \hline
         6.6880$\cdot10^{-2}$ & 1.5507$\cdot10^{-1}$ & - \\
         12.635 & 16.929 & - \\ \hline
    \end{tabular}
    \end{minipage}
    \caption{Oscillator parameters for all materials used in this study with references. Where necessary, the models were rewritten to fit the form of Eq.~(\ref{eq:osc}) and the parameters were transformed to eV units. Transformed values are given with five significant digits. Values in rad/s can be obtained  by multiplying with e/$\hbar$. The static dielectric value for bromobenzene is then manually set to 5.37\cite{ZWOL} as described in Sec.~\ref{sec:modeling_dielectric_functions}.}
    \label{tab:modells}
\end{table}

\end{document}